
{\magnification=\magstep1
\def\refs{\leftskip=.3truein\parindent=-.3truein}
\def\endrefs{\leftskip=-.3truein\parindent=.3truein}
\baselineskip=24pt
\centerline{STABILITY IN THE WEAK VARIATIONAL PRINCIPLE OF BAROTROPIC FLOWS}

\bigskip

\centerline{Asher Yahalom }

\centerline{The Racah Institue of Physics, Jerusalem 91904, Israel}

\vfill\eject

\centerline{\bf Abstract}
\noindent
	I find conditions under which the
 "Weak Energy Principle" of Katz, Inagaki and Yahalom (1993)
gives necessary and sufficient conditions. My conclusion is that,
 necessary and sufficient conditions of stability are obtained
when we have only two mode coupling in the
gyroscopic terms of the perturbed Lagrangian. To illustrate
the power of this new energy principle, I have calculated
 the stability limits of two
dimensional configurations such as ordinary Maclaurin disk, an infinite self
gravitating rotating sheet, and a two dimensional Rayleigh  flow
which has well known sufficient conditions of stability.
All perturbations considered are in the same plane as the configurations.
The limits of stability are identical with those
 given by a dynamical analysis when available, and with the
results of the strong energy principle analysis when given.
 Thus although the "Weak Energy" method is mathematically more simple
than the "Strong Energy" method of Katz, Inagaki and Yahalom )1993)
since it does not involve solving second order partial differential
 equations, it is by no means less effective.

\bigskip
Key words: Energy variational principle; Self-gravitating systems;
Stability of fluids.
\vfill\eject

\noindent
{\bf I. Introduction}
\bigskip
	The main purpose of this work is to study the stability
 features of weak variational principle of barotropic fluid dynamics
(Katz, Inagaki and Yahalom (1993) from now on paper I).
This principle has the advantage that all equations of barotropic
fluid dynamics are derived
from one Lagrangian, i.e. Euler equations and continuity equation.
And therefore stability analysis in this formalism
does not involve solving partial differential equations
like in the dynamical perturbation method or the strong variational
principle (see paper I).
 However, it has the disadvantage of lacking standard form.
This Lagrangian contain only a term depending on the
degrees of freedom (a potential) and a term linear in time derivatives
 (gyroscopic term), but lacks a term quadratic
 in time derivatives (kinetic term). The above fact is the cause of
some novel stability features I discuss below.

The plan of this paper is as follows, section II contains a quick
introduction to the weak variational principle of barotropic
flows  given in  (paper I),
 some modifications are introduced for both 3-D and 2-D cases .
 In section III I overview the stability theory of a system
which does not contain terms quadratic in time derivatives
(gyroscopic system), a comparison between stability predictions
using the strong variational principle and weak variational principle
is also given.

 In section IV I introduce the general form of the second variation of
a barotropic flow potentail. Later a few illustrations of
stability analysis using the weak energy principle
in two dimensional flows is given. In section V I give a general formalism
of uniformly rotating galactic flows. In section VI I study a specific case
of the above flows, the uniform rotating sheet which was first analyzed by
Binney J. \& Tremaine S. (1987). In section VII I analyse the stability of
another
 galactic model known as
the "Maclaurin disks" which also appears at Binney J. \& Tremaine S. (1987).
This model is analysed dynamicaly in (Binney J. \& Tremaine S. (1987))
by solving three equations (Euler equations in two dimensions and
the continuity equation). It is also analysed in (Yahalom, Katz \&
Inagaki 1994 from now on paper II) using the strong energy method and solving
one equation
(continuity equation). Here it is analysed  without solving
any equation, but merely diaganolizing the appropriate potential.
In section VIII I analyse  the two dimensional Rayleigh flows ,
where sufficient conditions of stabilty were found by Lord Rayleigh (1880).

\bigskip

\noindent
{\bf II. The Weak Variational Principle}

The weak variational principle is derived in paper I
from the Lagrangian (5.16) combined with (6.12) :
$$ L = \int [\vec w \cdot \vec v - \left({1 \over 2}
   \vec v^{2} + \varepsilon + {1 \over 2} \Phi\right)]\rho d^{3}x
 + \vec b \cdot (\vec P-\vec P_0) + \vec \Omega_c \cdot (\vec J -\vec J_0)
\eqno(II.1)$$
We use the following notations: for the positions of fluid
elements $ \vec r $ or $ (x^K)=(x,y,z);$   $K,L,... = 1,2,3$
the density of matter is $ \rho $;
 $ \vec v$ is the velocity field in inertial coordinates.  $ \vec b$ and
 $\vec \Omega_c $
serve as Lagrange multipliers with respect to linear and angular momenta
respectively
given by:
$$ \vec P  = \int \vec v \rho d^3 x  \eqno(II.2.a)$$
and
$$  \vec J = \int \vec r \times \vec v \rho d^3x  \eqno(II.2.b)$$
$\varepsilon (\rho)$ is the specific internal energy of the barotropic
fluid, related to the pressure and the specific enthalpy:
$$ \varepsilon(\rho) = h - {P \over \rho}  \qquad P = \rho^{2}{\partial
   \varepsilon\over \partial \rho}   \eqno(II.3)$$
$\Phi$ is the internal gravitational potential given by:
$$\Phi= - G \int {\rho(\vec r^{~'})\over{|\vec r - \vec r^{~'}| }}d^3 x'.
    \eqno(II.4)$$
In our formalism $ \vec v$ has a Clebsch form:
$$ \vec v = \alpha \vec \nabla \beta + \vec \nabla \nu. \eqno(II.5)$$
And $\rho$ is  given by:
$$ \rho =   {\partial (\alpha, \beta, \mu) \over \partial (x, y, z)}.
\eqno(II.6)$$
$\vec w$ is defined by:
$$ \dot \alpha + \vec w \cdot \vec \nabla \alpha = 0 \qquad
\dot \beta + \vec w \cdot \vec \nabla \beta = 0 \qquad
\dot \mu + \vec w \cdot \vec \nabla \mu = 0 .\eqno(II.7)$$
Hence our system is described by four trial functions $\alpha, \beta, \mu, \nu$
and two Lagrange multipliers $ \vec b$ and $\vec \Omega_c $. Taking the
variation
of Lagrangian (II.1) with respect to the four trial functions gives the Euler
and
mass conservation equations in moving coordinates. Taking the variation
of  Lagrangian (II.1) with respect to the two Lagrange multipliers gives the
fixation
of linear and angular momentum. Taking the variation of the potential part of
(II.1) gives the equation of a stationary barotropic fluid.
 Further details can be found in paper I.
I will know modify the Lagrangian slightly. Rewriting the first part of
equation (II.1)
using equation (II.5) :
$$ \int \vec w \cdot \vec v \rho d^{3}x = \int \vec w \cdot (\alpha \vec \nabla
\beta +
 \vec \nabla \nu) \rho d^{3}x
=\int (-\alpha \dot \beta \rho -\nu \vec \nabla \cdot (\rho \vec w)) d^{3}x
\eqno(II.8)$$
where in the second equality we integrated by parts and neglected boundary
terms
(we also assumed that $\nu$ is single valued) and used equation (II.7).
Thus we obtain:
$$ \int \vec w \cdot \vec v \rho d^{3}x =\int (-\alpha \dot \beta \rho + \nu
\dot \rho) d^{3}x
 =\int (-\alpha \dot \beta -\dot \nu)\rho d^{3}x  \eqno(II.9)$$
where in the second equality we added a full time derivative.
Inserting equation (II.9) into equation (II.1)  and using equation (II.5) we
obtain:
$$ L = \int \left[-\alpha \dot \beta -\dot \nu -
 {1 \over 2}  \vec (\alpha \vec \nabla \beta +\vec \nabla \nu)^{2}
+ \varepsilon + {1 \over 2} \Phi\right] \rho d^{3}x
 + \vec b \cdot (\vec P -\vec P_0) + \vec \Omega_c \cdot (\vec J -\vec J_0).
 \eqno(II.10)$$
In 2-D flows the formalism is modified slightly. Euler equations have only two
components,
 linear momentum has two components $\vec P$. Angular momenntum $J$ and
vorticity
$\omega = \vec \nabla \times \vec v \cdot \vec 1_z$ have one component.
The density $\rho$ is given by:
$$ \rho = \Sigma \delta_D (z) = \lambda(\alpha) {\partial (\alpha, \beta) \over
 \partial (x, y)} \delta_D (z). \eqno(II.11)$$
$ \delta_D$ is diracs delta, $\Sigma $ is the surface density,
and $\lambda$ is a function of $\alpha$ depending on the flow under
consideration.
 The internal energy $ \varepsilon $ and pressure $P$ are considred to be
functions of $\Sigma$.
The Lagrangian of a 2-D flow does not depend on $\mu$ and is given by:
$$ L = \int \left[-\alpha \dot \beta -\dot \nu - {1 \over 2}
   (\alpha \vec \nabla \beta +\vec \nabla \nu)^{2}
+ \varepsilon + {1 \over 2} \Phi\right]
\Sigma d^{2}x + \vec b \cdot (\vec P-\vec P_0) +  \Omega_c ( J - J_0).
\eqno(II.12)$$
For Further details see paper II.

\noindent
{\bf III. Stability of Gyroscopic Systems}

 It is our purpose now to derive from the functional form of the potential term
in
the Lagrangian (II.10) the conditions under which a given stationary
configuration
becomes stable or unstable. To do this we study a system with N degrees
of freedom that is described by the Lagrangian:
$$ L = G - V = b_i (q) \dot q^i - V(q) \eqno(III.1)$$
(a summation agreement is assumed) the energy will become
$$ E = V(q). \eqno(III.2)$$
This means that the system can propogate only on a manifold of
 equipotential, if the
equipotential
manifold is zero dimensional as in the case of a maximum or a
minimum of the potential
than the system is bound to
remain on this point i.e. we have a sufficient condition of stability. This can
also
be deduced from the following argument, suppose we distort our
equilibrium configuration given by  $q^i = q^i_0 ,
 \dot q^i_0 = 0$ and ${\partial V \over
\partial q_i} |_{q_0} = 0 $ slightly such that initial
 conditions and energy are slightly different. Since this is done around
equilibrium
 there is no first order contribution to the variation of energy and we obtain:
$$ \delta^2 E = {\partial^2 V \over \partial q_i \partial q_j}|_{q_0}
 \delta q_i \delta q_j \equiv V_{ij}  \delta q_i \delta q_j .\eqno(III.3)$$
 From equation (III.3) it is easy to see that if the matrix
$V_{ij} $ has a definite sign (i.e. we
have a minimum or a maximum) , the system is bounded (i.e. stable). Let us now
take
the dynamical point of view, writing down the linearized
Euler-Lagrange equations of this system we obtain:
$$ V_{ij} \delta q_i = ({\partial b_i
 \over \partial q_j} - {\partial b_j \over \partial q_i})|_{q_0} \delta \dot
q_i
 \equiv b_{ij} \delta \dot q_i. \eqno(III.4.a)$$
The above expression can be derived from the perturbed Lagrangian:
$$ \delta^2 L = \delta^2 G - \delta^2 V = b_{ij} \delta q_i
 \delta \dot q_j  -  V_{ij} \delta q_i  \delta q_j \eqno(III.4.b)$$
Now suppose that:
$$ \delta q_i  \propto e^{ \omega t} \eqno(III.5)$$
this means that:
$$ V_{ij} \delta q_i =\omega b_{ij} \delta q_i. \eqno(III.6)$$
in order to obtain the eigen frequencies we must solve the equation.
$$ det|V_{ij} - \omega  b_{ij}| = 0. \eqno(III.7)$$
 Assume without the loss of generality that $V_{ij}$ is diaganolized. Taking a
two dimensional perturbation (a two mode coupling in $\delta^2 G $) we have:
$$ \omega^2 = - {V_{11}  V_{22} \over b_{12}^2} \eqno(III.8)$$
in this case a definite sign is a necessary and sufficient condition of
stability as
can be clearly seen from equation (III.8). Moving on to the third dimensional
case:
$$ \omega^2 = - {V_{11}  V_{22} V_{33} \over b_{12}^2 V_{33} +b_{13}^2 V_{22} +
b_{23}^2 V_{11} } \eqno(III.9)$$
here a definite sign is not necessary, take for example:
$$ V_{11}<0 \   V_{22}<0 \  V_{33}>0 \qquad  {\rm with} \qquad
  b_{12}^2 V_{33} +b_{13}^2 V_{22} + b_{23}^2 V_{11}>0. \eqno(III.10)$$
Thus we conclude that a neseccary and sufficient condition of stability appear
only in
the case where the gyroscopic term couple only two modes, for more complex
coupling
the energy criterion is only sufficient.
\bigskip

\noindent
{\it III.1 Strong systems derived from Gyroscopic systems}

 In this section we diverge from the main topic of this paper in order
to study the relations between weak and strong variational principles
(paper I).
Observing closely the Lagrangian (II.10) we see that it contains two kinds of
variables,
i.e. we have $N+M$ variables $q_i,\nu_a$ (we take $i,j,k \in [1..N]$ and
$a,b,c \in [1..M]$) , the Lagrangian has  the following  structure:
$$L_w = b^1_i(q) \dot q_i + b^2_a (q) \dot \nu_a - V^w (q,\nu), \qquad V^w =
V^0(q)+V^1_a(q)
\nu_a+{1 \over 2}
\nu_a \nu_a \eqno(III.11)$$
The equations of motion of this system are:
$$b^1_{[i,k]} \dot q_k - b^2_{a,i}\dot \nu_a = -V^w_{,i} \eqno(III.12.a)$$
$$\nu_a = -b^2_{a,i} \dot q_i-V^1_a. \eqno(III.12.b)$$
Using equation (III.12.b) we eliminate $\nu_a$ in (III.11) and obtain:
$$L_s = {1 \over 2} g_{ij} \dot q_i \dot q_j +b^s_i \dot q_i -V^s, \qquad
g_{ij} =
 b^2_{a,i} b^2_{a,j}, \qquad
b^s_i = b^1_i +V^1_a b^2_{a,i}, \qquad V^s=V^0 -{1 \over 2} V^1_a V^1_a.
\eqno(III.13)$$

The euilibriums of the two systems are the same.
 In equilibrium $\nu_{a0} =-V^1_a$ (see (III.12.b)),
inserting this into equation (III.12.a) we obtain:
$$ V^w_{,i} = (V^0 -{1 \over 2} V^1_a V^1_a)_{,i} = V^s_{,i} = 0.
\eqno(III.14)$$
Without loss of generality we consider the stability of an equilibrium given by
$\nu_{a0} = 0,q_{i0}$, where $q_{i0}$ are defined by equation (III.14).

 The energy criterion of the two systems is also the same:
For a quadratic system such as $L_s$ the energy criterion takes the
 form $V^s_{,ij}|_0 \delta q_i \delta q_j > 0$,
while for a gyroscopic system such as $L_w$ we demand that
 $\Delta^2 V^w = (\delta \nu_a + V^1_{a,i} \delta q_i)^2
+ (V^0_{,ij} - V^1_{a,i} V^1_{a,j}) \delta q_i \delta q_j =
 (\delta \nu_a + V^1_{a,i} \delta q_i)^2 +
V^s_{,ij}|_0 \delta q_i \delta q_j$ has a definite sign,
i.e. $V^s_{,ij}|_0 \delta q_i \delta q_j > 0$.

It is not to be understood that transforming the weak
 Lagrangian into $L_s$ has no benefit from the
point of view of stability analysis. In fact the strong
 principle provides us with necessary and sufficient
conditions beyond what we obtain from (III.8).
 Every perturbation of $L_s$ implicitly includes all $\delta \nu$
perturbations, that is we obtain necessary and sufficient
 conditions for more general perturbations than
are obtained in the two mode analysis of equation (III.8).
 For example for a single mode, say $\delta q_1$  of $L_s$
the gyroscopic term vanishes and we have a
 necessary and sufficient condition of stability:
$$ V^s_{,11} = V^0_{,11}-V^1_{a,1} V^1_{a,1} > 0 \eqno(III.15.a)$$
this can be compared to a two mode analysis of $L_w$, with $\delta q_1$
and $\delta \nu_2$ for which we obtain:
$$  V^0_{,11}-V^1_{2,1} V^1_{2,1} > 0. \eqno(III.15.b)$$
 Thus, necessary and sufficient conditions obtained from $L_s$ are more strict.
 However, in $L_s$ one
excludes initialy some of the less general perturbations of
$\delta \nu$ in this sense analysis through $L_s$
misses some of the necessary and sufficient conditions available in the
dynamics.
 We conclude that $L_s$ and $L_w$ are complementary ways for obtaining
 necessary and sufficient conditions of stability, for
sufficient condition the methods give the same results.
\bigskip

\noindent
{\bf IV. The Energy Criteria for Stability of 2-D Flows }

\noindent
{\it IV.1  Basic Identities of Perturbation Theory}

In order to obtain the concrete barotropic flow form of $\delta^2 V$ and
$\delta^2 G$
of equations (III.4.b) that are needed for stability analysis, a few basic
identities
and notations will be given here, this will make our future calculations
easier.

We look at the following Eulerian small displacements:
$$\alpha_0 (x,y) \rightarrow  \alpha_0 (x,y) + \delta  \alpha (x,y), \
\beta_0 (x,y) \rightarrow  \beta_0 (x,y) + \delta  \beta (x,y), \
 \nu_0 (x,y) \rightarrow  \nu_0 (x,y) + \delta  \nu (x,y) \eqno(IV.1.a)$$
and also:
$$ \Omega_{c0} \rightarrow \Omega_{c0} + \delta \Omega_c, \qquad
 \vec b_0 \rightarrow  \vec b_0 + \delta \vec b. \eqno(IV.1.b)$$
The subscript $_0$ denotes stationary quantites.
It is convenient to introduce the often used $\vec \xi$,
which is defined by:
$$ \alpha (\vec R + \vec \xi) + \delta \alpha (\vec R + \vec \xi)
 \equiv \alpha (\vec R) \quad
\beta (\vec R + \vec \xi) + \delta \beta (\vec R + \vec \xi)
 \equiv \beta (\vec R)\eqno(IV.2.a)$$
to order 1,
$$ \delta \alpha = - \vec \xi \cdot \vec \nabla \alpha, \quad
\delta \beta = - \vec \xi \cdot \vec \nabla \beta. \eqno(IV.2.b)$$
Having defined $\vec \xi$, we can also define the Lagrangian displacement
$\Delta$:
$$ \Delta \equiv \delta + \vec \xi \cdot \vec \nabla \eqno(IV.3)$$
And ofcourse $\Delta \alpha =  \Delta \beta = 0 $. The following identities
will
serve us at future calculations:
$$ \Delta \vec R = \vec \xi, \qquad
\delta \vec \nabla = \vec \nabla \delta, \qquad
\Delta \vec \nabla  = \vec \nabla \Delta  - \vec \nabla \vec \xi \cdot \vec
\nabla,
 \qquad \Delta \vec \xi = 0.  \eqno(IV.4)$$

The variations of both the surface density $\Sigma$ and the velocity $\vec v$
can be derived from equations (II.5) and (II.11) (for details see paper I
 see also paper II):
 $$ \Delta \Sigma = - \Sigma_0 \vec \nabla \cdot \vec \xi, \qquad
\delta \Sigma = - \vec \nabla \cdot (\Sigma_0 \vec \xi) \eqno(IV.5.a)$$
$$\Delta \vec v = -\vec \nabla \vec \xi \cdot \vec v_0 +
 \vec \nabla \Delta \nu, \qquad \delta \vec v = -\vec \nabla \vec \xi \cdot
\vec v_0
+ \vec \nabla \Delta \nu - \vec \xi \cdot \vec \nabla \vec v_0, \eqno(IV.5.b)$$
we shall also need the second variation of velocity:
$$ \Delta^2 \vec v = -2 \vec \nabla \vec \xi \cdot \Delta\vec v
+ \vec \nabla \Delta^2 \nu . \eqno(IV.5.c) $$

\noindent
{ \it IV.2 The Variation of The Potential $V$ }

We now give the concrete expression of (III.3)
for the potential of a barotropic~~flow.
The potential part of the Lagrangian (II.12) is:
$$ V = \int \left[{1 \over 2} \vec v^{2} + \varepsilon(\Sigma) +
 {1 \over 2} \Phi \right] \Sigma d^{2}x - \vec b \cdot (\vec P-\vec P_0)
-  \Omega_c ( J - J_0).  \eqno(IV.6)$$
Notice that this is also the energy according to equation (III.2).

The condition that $\Delta V = 0$ for arbitrary perturbations of the form
(IV.1) is that Euler and the continuity equations of
stationary motion are satisfied in moving coordinates
as well as the fixation of linear and angular momentum:
$$(\vec v_0-\vec b_0 - \Omega_{c0} \vec 1_z \times \vec R) \cdot \vec
\nabla\vec v_0
 + \Omega_{c0} \vec 1_z \times \vec v_0 + \vec \nabla (h_0 + \Phi_0) = 0,
 \eqno(IV.7.a)$$
and also:
$$\vec \nabla \cdot (\Sigma_0 (\vec v_0-\vec b_0 - \Omega_{c0} \vec 1_z
\times \vec R)) = 0, \qquad J = J_0, \qquad \vec P =\vec P_0. \eqno(IV.7.b)$$
For details see paper I.

We now look at the second variation $\delta^2 V = \Delta^2 V$
at a configuration satisfying equations (IV.7):
$$\Delta^2 V = \int \{ (\Delta \vec v)^2 + \vec v \cdot \Delta^2 \vec v
 + \Delta [\vec \xi \cdot \vec \nabla (h + \Phi)] \} |_0 \Sigma_0 d^2 x
- 2 \Delta \vec b \cdot \Delta \vec P -2 \Delta \Omega_c \Delta J
- \vec b_0 \cdot \Delta^2 \vec P - \Omega_{c0} \Delta^2 J,\eqno(IV.8.a)$$
where:
$$ \Delta \vec P = \int \Delta \vec v \Sigma_0 d^2 x , \qquad
\Delta^2 \vec P = \int \Delta^2 \vec v \Sigma_0 d^2 x , \eqno(IV.8.b)$$
and:
$$ \Delta J = \vec 1_z \cdot \int (\vec R \times \Delta \vec v +
\vec \xi \times \vec v_0 ) \Sigma_0 d^2 x , \qquad
\Delta^2 J = \vec 1_z \cdot \int ( 2 \vec \xi \times \Delta \vec v +
\vec R \times \Delta^2 \vec v ) \Sigma_0 d^2 x . \eqno(IV.8.c)$$

We demand now that:
$$ \Delta J =  \Delta \vec P  = 0 \eqno(IV.9)$$
this will constrain the perturbed configuration to have the same
linear and angular momentum as the stationary one and will define
hopefully both $\Delta \vec b$ and $\Delta \Omega_c$. The perturbed
potential $\Delta^2 V$ will now have the form:
$$\Delta^2 V = \int \{ (\Delta \vec v)^2 + \vec v \cdot \Delta^2 \vec v
 + \Delta [\vec \xi \cdot \vec \nabla (h + \Phi)] \} |_0 \Sigma_0 d^2 x
 - \vec b_0 \cdot \Delta^2 \vec P - \Omega_{c0} \Delta^2 J,\eqno(IV.10)$$
In configuration having special symmetry $\Omega_{c0}$ or $\vec b_0$ or both,
may not be defined by equation (IV.7.b), in this case we will demand:
 $\Delta^2 J =0 $ or $\Delta^2 \vec P =0 $ or both.
In order to see whether the bilinear form $\Delta^2 V$ is positive around
a certain stationary configuration we will have to diagonolize it first
this will be done for a few examples in later parts of this paper.

\noindent
{ \it IV.3 The Variation of The Kinetic Term $G$ }

Although sufficient conditions of stability can be derived from expression
 (IV.10) by checking its positivity for certain perturbations, one cannot
claim any new knowledge of the stability of the system, unless the perturbation
is the most general one. The reason for this is that if $\Delta^2 V > 0$ it may
yet be negative for another form of perturbation and hence stability cannot
be claimed. If on the other hand $\Delta^2 V < 0$ the system is not unstable
 because as we emphasized the $\Delta^2 V > 0$ condition is only sufficient.
Thus, in order to achive new information we must check how the perturbation
under
consideration appears at the kinetic term $\delta^2 G = \Delta^2 G$.
As we saw in section III
 we will have also a necessary condition if the perturbation is of the double
mode type. The term G of the Lagrangian (II.12) is of the form:
$$ G = \int \left[-\alpha \dot \beta -\dot \nu \right] \Sigma d^2 x
\eqno(IV.11)$$
The first  variation of this expression is:
$$ \Delta G = \int \left[-\alpha \Delta \dot \beta - \Delta \dot \nu \right]
\Sigma d^2 x ,\eqno(IV.12)$$
notice that $ \Delta \dot f = \dot (\Delta f) - \dot {\vec \xi} \cdot \vec
\nabla
f $ hence:
$$ \Delta G = \int \left[\dot {\vec \xi} \cdot  \vec v - \dot {(\Delta \nu)}
\right]
\Sigma d^2 x.\eqno(IV.13)$$
For the second variation near equlibrium we obtain:
$$ \Delta^2 G = \int \left[\dot {\vec \xi} \cdot \Delta \vec v +
\Delta \dot {\vec \xi} \cdot \vec v - \Delta \dot {(\Delta \nu)} \right]
\Sigma_0 d^2 x.\eqno(IV.14)$$
Using equation (IV.5.b) we obtain:
$$ \Delta^2 G =  \int \left[2 \dot {\vec \xi} \cdot \Delta \vec v + \dot
{(\Delta^2
 \nu)}\right] \Sigma_0 d^2 x.\eqno(IV.15)$$
The last term is a full time derivative and does not contribute to the
equations of
motion and thus can be neglected, finally we obtain:
$$ \Delta^2 G =  \int \left[2 \dot {\vec \xi} \cdot \Delta \vec v \right]
\Sigma_0 d^2 x.\eqno(IV.16)$$
\vfill\eject

\noindent
{\bf V. Applications to Galactic Flows of Uniform Rotation  }

\noindent
{\it V.1. Stationary Configurations }

	The following is taken from Binney and Tremaine (1987) with some
changes of notations. We consider disks of fluid rotating with uniform angular
 velocity $ \vec   \Omega = \vec 1_z \Omega $ and the velocity field in
inertial
 coordinates is thus:
$$ \vec v_0 = \vec \Omega \times \vec R = \Omega R^2 \vec \nabla \varphi
\eqno(V.1)$$
 $ \varphi $ is the polar angle, $R$ the radial distance. And the density is:
$$ \Sigma_0 = \Sigma_C \tilde \Sigma_0 (R) \eqno(V.2)$$
Inserting equations (V.1) and (V.2) into (IV.7) (taking into account the
symmetry of
the potential and the enthalpy derived) we see that
the variable $\Omega_{c0}$ is not defined due to the symmetry of
the problem and also:
$$ \vec b_0 = 0 \eqno(V.3)$$
Pressure and specific enthalpy are given by arbitary equations of state:
$$ P =  P(\Sigma), \quad h = h(\Sigma). \eqno(V.4)$$

\noindent
{\it V.2. Global Properties of Perturbed Configurations. }

We now look at the first order perturbation of angular momentum $\Delta J$ of
the above configurations as it appears in equation (IV.8.c).
We calculate the first term of the above equation and use both equations (V.1)
 and (IV.5.b):
$$ \int \vec R \times \Delta \vec v|_0 \Sigma_0 d^2 x =
\int \vec R \times (-\vec \nabla \vec \xi \cdot \vec v_0+\vec \nabla \Delta
\nu)
 \Sigma_0 d^2 x =  - \Omega \int \vec \xi \cdot \vec R \Sigma_0 d^2 x.
\eqno(V.5)$$
 Calculating the second term we obtain:
$$ \int \vec \xi \times \vec v_0 \Sigma_0 d^2 x =
\int \vec \xi \times (\vec \Omega \times \vec R) \Sigma_0 d^2 x =
\Omega \int \vec \xi \cdot \vec R \Sigma_0 d^2 x \eqno(V.6)$$
Since $(V.5) + (V.6) \equiv 0$, we derive that
$$ \Delta J \equiv 0 \eqno(V.7)$$
Since $\Omega_c$ is not defined
 we must demand $ \Delta^2 J = 0$.
Taking $ \Delta^2 J= 0$ we are able to obtian differential equality convenient
 for calculating $\Delta^2 V$. It is obtained from (A.8) in appendix A
$$ \Delta^2 V  = \int \{  \Sigma_0 (\delta \vec v)^2 +
 \delta \Sigma \delta (h + \Phi) \} d^2 x \eqno(V.8)$$
\bigskip

\noindent
{\it V.3. $\vec \xi$ defined by scalar functions}

	 It will appear very convinient to define $ \vec \xi$
in terms of two independant non dimensional infinitesimal
 scalars $ \eta$ and $ \psi$ as follows:
$$ \vec \xi = a^2 [\vec \nabla \eta + rot \vec \psi], \qquad \vec \psi = \vec
1_z \psi.
 \eqno(V.9)$$
$a$ is a scale to be defined later.
 $\eta$ is thus defined in terms of $\vec \xi$ by
$$ \triangle \eta = {1 \over a^2} \vec \nabla \cdot \vec \xi \eqno(V.10)$$
Some boundary conditions are needed to make $\eta$ unique. If we take, say,
$$\eta|_B = 0 \eqno(V.11)$$
equation (V.10) has a unique solution. Equation (V.10) represents also the
condition of
integrability of (V.9), considered as a set of two first order differential
equations
 for $\psi$, given $\vec \xi$ and $\eta$. Thus the $\psi$ equations  are
integrable and
 define $\psi$ up to a constant.

	Following equation (IV.5.b) and equations (V.1) and (V.9), $\delta \vec v$ and
$\Delta \vec v$ may be
 written as a gradient plus a rotational, always a convenient form for vector
fields:
$$ \delta \vec v = a^2 \Omega [ \vec \nabla \zeta+ rot (2 \eta \vec 1_z)]
\eqno(V.12.a)$$
$$ \Delta \vec v = \Omega [\vec \nabla (\zeta + \psi) + rot ( \eta \vec 1_z)]
 \eqno(V.12.b) $$
in which
$$ \zeta= {1 \over a^2 \Omega}(\Delta \nu - \vec \xi \cdot \vec v_0)- 2\psi
 \eqno(V.13).$$
\bigskip

\noindent
{\bf VI. Application to a Uniform Rotating Sheet }
\bigskip

\noindent
{\it VI.1. Stationary Configurations }

In this section we study a specific case of the flows described in section V.
We consider an infinite uniform rotating sheet.
Since the density is uniform we have:
$$ \tilde \Sigma = 1 \eqno(VI.1)$$
There is no self gravitational field in the plane of the sheet, however, to
maintain
equilibrium we assume the existance of a radial symmetrical external potential:
$$ \Phi_0 = \Phi_0(R) \eqno(VI.2)$$
Euler's equations (IV.7.a) relates $\Omega$ to $\Phi_0(R)$:
$$  {\partial \Phi_0 \over \partial R} = \Omega^2 R \eqno(VI.3)$$
\vfill\eject

\noindent
{\it VI.2. Fourier Decompositions of  $\vec \xi$ , $\delta \Sigma$, $\delta h$,
$\delta \Phi$ and $\zeta$}

	In order to diagonolize the second order perturbation of the potential (V.8)
 around our present stationary configuration, we will
write our perturbed quantities as fourier transforms.
 	First we write $\eta$ and $ \psi$  as a fourier transform:
$$ \eta (\vec R) = {1 \over 2 \pi} \int_{-\infty}^{\infty} \eta (\vec k)  e^{i
\vec k
\cdot \vec R} d^2 k  \eqno(VI.4.a)$$
$$ \psi (\vec R) = {1 \over 2 \pi} \int_{-\infty}^{\infty} \psi (\vec k)  e^{i
\vec k
\cdot \vec R} d^2 k.  \eqno(VI.4.b)$$

 	 Next we calculate the Fourier decomposition of
 $\delta \Sigma$. Using equations (IV.5.a) and
 equation (V.9), taking the scale $a=1$, $\delta \Sigma$ can be written as:
$$ \delta \Sigma = - \Sigma_C \triangle \eta  \eqno(VI.5)$$
and from equations (VI.4) we find that $ \delta \Sigma $ has the following
 expansion:
$$ \delta \Sigma = {1 \over 2 \pi} \int_{-\infty}^{\infty} \delta \Sigma (\vec
k)
 e^{i \vec k \cdot \vec R} d^2 k  \eqno(VI.5.a)$$
in which
$$ \delta \Sigma (\vec k) =  \Sigma_C k^2 \eta (\vec k) \eqno(VI.5.b)$$
With the expansion of $\delta \Sigma$ we obtain directly the Fourier
decomposition of
 $\delta h$:
$$ \delta h = {\partial h \over \partial \Sigma} \delta \Sigma \equiv  {v_s^2
\over
\Sigma_C} {1 \over 2 \pi} \int_{-\infty}^{\infty} \delta \Sigma (\vec k)
 e^{i \vec k \cdot \vec R} d^2 k  \eqno(VI.6)$$
where $v_s$ is the sound velocity.
{}From the varied Poisson equation we calculate the perturbed gravitational
potential:
$$ \triangle \delta \Phi = 4 \pi G \delta \Sigma \delta_D (z) \eqno(VI.7)$$
where $\delta_D (z)$ is the Dirac function, we find the solution $\delta \Phi$
which
 has been given by Binney \& Tremaine (1987):
$$ \delta \Phi = -\int_{-\infty}^{\infty} \delta \Sigma (\vec k){ 2 \pi G \over
|k|}
 e^{i \vec k \cdot \vec R} d^2 k. \eqno(VI.8)$$
The only quantity which we nead to decompose in order to obtain the
expansion of $\delta \vec v$ is $\zeta$ which we now write as:
$$  \zeta = {1 \over 2 \pi} \int_{-\infty}^{\infty} \zeta (\vec k)
 e^{i \vec k \cdot \vec R} d^2 k. \eqno(VI.9)$$

\noindent
{\it VI.3. Global Constraints}

According to equation (IV.9) we must demand that $\Delta \vec P = \Delta J =
0$.
However, $\Delta J = 0$ for all possible perturbations as we derived in
equation (V.7).
Thus, we must take care only of $ \Delta \vec P $. In addition, due to the
symmetry of
the problem we saw in section V.2 that we must also demand $\Delta^2 J = 0$ .
replacing equation (V.12.b) into (IV.8.b),  we obtain:
$$ \Delta  \vec P = \Sigma_C \Omega \int \vec \nabla (\zeta + \psi) +
 rot ( \eta \vec 1_z) d^2 x = 0 \eqno(VI.10)$$
using stokes theorem this can be written as:
$$ \Delta  \vec P = \Sigma_C \Omega \oint [(\zeta + \psi) \vec 1_z \times d
\vec R +
\eta \vec 1_z \cdot d \vec R] = \Sigma_C \Omega \oint (\zeta + \psi) \vec 1_z
\times d
 \vec R = 0. \eqno(VI.11)$$
In order to fix the linear momenta we take both $\psi$ and $\zeta$ to be zero
at
infinity.
$$ \psi = 0 , \qquad \zeta = 0  \qquad R \rightarrow \infty \eqno(VI.12)$$
The constraint  $ \Delta^2 J =0$ becomes, after
inserting equations (V.12), (VI.5) and (IV.5.c) into
the second part of equation (IV.8.c):
$$ \Delta^2 J = -2 \Sigma_C \Omega \int \zeta \triangle {\partial \eta \over
\partial \varphi} d^2 x = 0 \eqno(VI.13)$$
This constraint  can be satisfied by taking $\eta (\vec k) = \eta(k)$
this will not modify the stability condition which which will obtain in the
next section since it depends only on $k$.

\noindent
{\it VI.4. Necessary and Sufficient Conditions for Stability of Infinite
Sheets}

	$\Delta^2 V$ is given by (V.8) near the stationary configuration.
To obtain the Fourier decomposition of $\Delta^2 V$, we replace
$\delta \Sigma$ , $\delta h$, $\delta \Phi$ and $\delta \vec v$ by their
respective
expansions (VI.5.b), (VI.6), (VI.8) and (VI.9) with (V.12.b),
 in $\Delta^2 V$ taking account of constraints
(V.11), (VI.12) and integrating we obtain the following results:
$$ \Delta^2 V= 2 \Sigma_C \int k^2 [\Omega^2 |\zeta|^2 +(4 \Omega^2 + k^2
v_s^2 - 2 \pi G \Sigma_C |k|) |\eta|^2]  d^2 k \eqno(VI.14)$$
Following section IV.2 the inequality $\Delta^2 V > 0$, calculable from (VI.14)
 gives sufficient conditions of stability. From (VI.14) we see that $\Delta^2 $
is positive for all $k$'s such that:
$$4 \Omega^2 + k^2 v_s^2 - 2 \pi G \Sigma_C |k| > 0 \eqno(VI.15.a)$$
this means that the sheet is stable to any wave-length if:
$${v_s \Omega \over G \Sigma_C} \ge {\pi \over 2} .\eqno(VI.15.b)$$
 The condition (VI.15) becomes sufficient and necessary when the gyroscopic
term for
 dynamical perturbations $\Delta^2 G$ couples only pairs of modes (see section
IV.3).

We can make a Fourier decomposition of $\Delta^2 G$, using  $\vec \xi$ given
 by (V.9) and and $\Delta \vec v$ given by (V.12.b), and using expansions
(VI.4)
and (VI.9):
$$ \Delta^2 G = 2 \Sigma_C \Omega  \int k^2 \zeta \dot \eta^* d^2 k +
 complex \  conjugate. \eqno(VI.16)$$
Necessary and sufficient conditions of stability are obtained
 from (VI.15) since $\Delta^2 G $ couples only pairs of
modes of $\eta$ and $\zeta$ . The condition (VI.15) was previously discovered
by
 Binney \& Tremaine (1987) using the dynamical perturbation method, this was
done by our
method without solving a single differential equation.
\bigskip

\noindent
{\bf VII. Applications to Maclaurin Disk }

\noindent
{\it VII.1. Stationary Configurations }

In this section we study a different case of the flows described in section V.
This time we study the stability of of the
Maclaurin disk which is a finite uniform rotating disk.
The density of this disk is given by:
$$ \tilde \Sigma = \sqrt {1 - {R^2 \over R_D^2}} \equiv \chi \qquad R \le R_D,
\eqno(VII.1.a)$$
$$ \tilde \Sigma = 0  \qquad \qquad R \ge R_D. \eqno(VII.1.b)$$
$R_D$ is the radius of the disk. We shall take $a=R_D$.
The self gravitational field in the plane of the disk is:
$$ \Phi_0 = {1 \over 2} \Omega_0^2 R^2 + const., \qquad \Omega_0^2 = {\pi^2 G
\sigma_C \over 2a}
 \eqno(VII.2)$$
$\Omega_0$ is the angular velocity of a test particle on a circular orbit.
See Binney and Tremaine (1987) for further details.
Here we allow only a certain equation of state  such that the Pressure and
specific enthalpy
 are given by:
$$ P =  \kappa \Sigma^3 \eqno(VII.3.a)$$
$$ h = {3 \over 2} \kappa \Sigma^2 \eqno(VII.3.b)$$
Euler's equations relate $\Omega$ to $\Omega_0$, $\Sigma_C$ and {\it a}:
$$ \Omega^2 = \Omega_0^2 - {3 \kappa \Sigma_C^2 \over a^2} \eqno(VII.4)$$

\noindent
{\it VII.2. Spherical Harmonic Decompositions of  $\vec \xi$ , $\delta \Sigma$,
$\delta h$,
  $\delta \Phi$ and $\zeta$}

  	We now decompose $\eta$ and $ \psi$  in normalised spherical harmonic
functions
 of $\chi$ defined in equation (VII.1) and $\varphi$.
We have:
$$ \eta = \sum_{l=m}^{\infty} \sum_{m=0}^{\infty} \eta_{lm} {\cal P}_l^m (\chi)
 e^{im \varphi} + c.c. \eqno(VII.5.a)$$
$$ \psi = \sum_{l=m}^{\infty} \sum_{m=0}^{\infty} \psi_{lm} {\cal P}_l^m (\chi)
 e^{im \varphi} + c.c. \eqno(VII.5.b)$$
 $\eta_{lm}$ and $\psi_{lm}$ are arbitary, infinitesimal, complex numbers.
 Associated Legendre polynomials are defined in the range
 $ -1 \le \chi \le 1 $, but $\eta$ and $ \psi$ are only defined in the range
 $ 0 \le \chi \le 1 $. We may extend $\eta$, $\psi$ to the negative region of
$\chi$ in
 any way we want. However, these functions have bounded gradients at $\chi =
0$.
 Indeed, since
$$ |\partial_R  \eta| < \infty,\qquad |\partial_R  \psi| < \infty
\eqno(VII.6.a)$$
$$ |{1 \over R} \partial_{\varphi} \eta| < \infty, \qquad |{1 \over R}
 \partial_{\varphi} \psi| < \infty \eqno(VII.6.b)$$
hold and since
$$ \partial_R  = - {1 \over a \chi} \sqrt{1- \chi^2} \partial_{\chi},
\eqno(VII.7)$$
then,
$$ \partial_{\chi} \eta|_{\chi=0} = \partial_{\chi}  \psi|_{\chi=0} = 0
.\eqno(VII.8)$$
Equations (VII.8) will be satisfied if we make symmetrical continuous
extensions
$$ \eta(\chi) = \eta(-\chi), \eqno(VII.9.a)$$
$$ \psi(\chi) = \psi(-\chi), \eqno(VII.9.b)$$
whose expansions are given by (VII.5) with $(l-m)$ even. The expansions in the
domain
$ 0 \le \chi \le 1 $, of {\it arbitrary} continuous $\eta$ and $\psi$  that
satisfy
(VII.6) everywhere are then given by (VII.5) with $(l-m)$ even [Arfken (1985)]
{}.
 The boundary conditions (V.11) gives the following relations among the
 $\eta_{lm}$'s: for every $m \ge 0$:
$$ \sum_{l=m}^{\infty}  \eta_{lm} {\cal P}_l^m (0) = 0, \qquad l-m \ \ even
\eqno(VII.10)$$
 Equations (VII.10) define, say, $\eta_{ll}$ in terms of $\eta_{l,m \ne l}$.

 	 We can now calculate the spherical harmonic decomposition of
 $\delta \Sigma$ using equations (IV.5.a) and  (V.9):
$$ \delta \Sigma = -a^2 \Sigma_C [ \vec \nabla \cdot (\chi \vec \nabla \eta)
+ \vec \nabla \chi \times \vec \nabla  \psi]. \eqno(VII.11)$$
and from equation (VII.11) and (VII.5) we find that $ \delta \Sigma $ has the
following
 expansion:
$$ \delta \Sigma = \Sigma_C \sum_{l=m}^{\infty} \sum_{m=0}^{\infty} \Sigma_{lm}
{{\cal P}_l^m (\chi) \over \chi} e^{im \varphi} + c.c. \qquad l-m \ even
\eqno(VII.12.a)$$
in which
$$ \Sigma_{lm} = [l(l+1) -m^2] \eta_{lm} + im \psi_{lm} \equiv k_{lm} \eta_{lm}
+ im
 \psi_{lm} \eqno(VII.12.b)$$
With the expansion of $\delta \Sigma$ we obtain directly through (VII.3.b) the
spherical
 harmonic decomposition of $\delta h$:
$$ \delta h = 3 \kappa \Sigma_C^2 \sum_{l=m}^{\infty} \sum_{m=0}^{\infty}
\Sigma_{lm}
 {\cal P}_l^m (\chi) e^{im \varphi} + c.c. \qquad l-m \ even \eqno(VII.13)$$
and from the varied Poisson equation :
$$ \triangle \delta \Phi = 4 \pi G \delta \Sigma \delta_D (z) \eqno(VII.14)$$
where $\delta_D (z)$ is the Dirac function, we find the solution $\delta \Phi$
which
 has been given by Hunter (1963):
$$ \delta \Phi = \Omega_0^2 a^2 \sum_{l=m}^{\infty} \sum_{m=0}^{\infty}
\Phi_{lm}
{\cal P}_l^m (\chi) e^{im \varphi} + c.c. \qquad l-m \ even \eqno(VII.15.a)$$
in which
$$ \Phi_{lm} = - g_{lm} \Sigma_{lm} \eqno(VII.15.b)$$
where
$$  g_{lm} = {(l+m)!(l-m)! \over 2^{2l-1} [({l+m \over 2})!({l-m \over 2})!]^2}
< 1
 \qquad l>0, \qquad l-m \ even \eqno(VII.15.c)$$
We now expand $\zeta$ in spherical harmonics:
$$  \zeta = \sum_{l=m}^{\infty} \sum_{m=0}^{\infty} \zeta_{lm} {\cal P}_l^m
(\chi)
e^{im \varphi} + c.c.,\qquad  l-m \ even.\eqno(VII.16)$$
\bigskip

\noindent
{\it VII.3. Global Constraints}

In this section as in VI.3 we must take care of $\Delta \vec P = 0$ and
$\Delta^2 J = 0$ .
Inserting equation (V.12.b) into the first expression of (IV.8.b) and using
expansions (VII.5) and (VII.16) we obtain the result:
$$ \zeta_{11}  = -\psi_{11} -i \eta_{11}.\eqno(VII.17)$$
i.e. in order to fix the momentum we must take $ \zeta_{11}$ to be a dependent
variable.
The constraint  $ \Delta^2 J =0$ becomes,
 (the calculation is in appendix B):
$$\eqalignno{& \Delta^2 J = 4 \pi \Sigma_C a^4 \Omega \sum m[  i \zeta_{lm}
\Sigma*_{lm} + m(|\psi_{lm}|^2 -|\eta_{lm}|^2)] +c.c. = 0 &(VII.18) \cr}$$
\vfill\eject

\noindent
{\it VII.4. Necessary and Sufficient Conditions for Stability of Maclaurin
Disks}

	In order to study the stability features of the Maclaurin Disks
we shall have to diagonolize $\Delta^2 V$  given by (V.8).
To obtain the spherical harmonic decomposition of $\Delta^2 V$, we replace
$\delta \Sigma$ , $\delta h$, $\delta \Phi$ by their respective
expansions (VII.12), (VII.13)  and (VII.15). Also we write $\delta \vec v$
by the represantation (V.12.a) using the expansions (VII.5.a), (VII.16) .
 The result is as  the follows [detailed calculations are
 given in appendix C]:
$$ \eqalignno{& \Delta^2 V=4\pi\Sigma_C \Omega^2 a^4 \sum_{(l=m)>1,
m=0}^{\infty} 4
 (k_{lm} - {m^2 \over k_{lm}}) |\eta_{lm}|^2 + k_{lm} |\zeta_{lm} +{2 im \over
k_{lm}} \eta_{lm}|^2 + \cr &  [- 1 + (1 - g_{lm}){\Omega_0^2 \over \Omega^2}]
|\Sigma_{lm} |^2  &(VII.19) \cr}$$
where we have taken into account the constraint (VII.17).
Following section IV.2 the inequality $\Delta^2 V > 0$, calculable from (VI.19)
with
the constraint (VII.18), gives sufficient conditions of stability.
 $\Delta^2 V $ has the form
 $ { \Omega_0^2 \over \Omega^2} A^2-B > 0$. If $B < 0$, the disk is stable for
any
 ${ \Omega^2 \over \Omega_0^2}$. For instance, Maclaurin disks are stable to
any
 perturbation that keeps the same densities at the displaced points
 $(\delta \Sigma = 0)$. We are naturally interested in perturbations that might
upset
 stability and for which $B > 0$. If $B > 0$ then ${ \Omega^2 \over
\Omega_0^2}$ must
 satisfy the following inequality
$$ \eqalignno{& Q \equiv { \Omega^2 \over \Omega_0^2} <
{\sum_{(l=m)>1,m=0}^{\infty}
 (1 - g_{lm}) |\Sigma_{lm} |^2 \over \sum_{(l=m)>1,m=0}^{\infty}
 |\Sigma_{lm} |^2 - 4 (k_{lm} - {m^2 \over k_{lm}}) |\eta_{lm}|^2 -
 k_{lm} |\zeta_{lm} +{2 im \over k_{lm}} \eta_{lm}|^2 } \cr & &(VII.20) \cr}$$
 The condition (VII.20) supplied by the constraint (VII.18)
becomes sufficient and necessary when the gyroscopic term for
 dynamical perturbations $\Delta^2 G$ couple only two modes (see section III).
 It is therefore important to calculate $\Delta^2 G$.
We can make a spherical harmonic decomposition of $\Delta^2 G$ given in
equation (IV.16).
A straightforward subtitution of $\vec \xi$ given
 by equations (V.9) and (VII.5) and $\Delta \vec v$ given by equations
(V.12.b), (VII.5) and
 (VII.16) leads to the following expression for $\Delta^2 G$:
$$ \Delta^2 G = 4  \pi \Sigma_C \Omega a^4 \sum_{l,m
\ne 1,1} [\dot \Sigma_{lm} \zeta_{lm}^* + im( \dot \eta_{lm}^* \eta_{lm} - \dot
\psi_{lm}^* \psi_{lm})] + complex \  conjugate \eqno(VII.21)$$
in which (VII.17) and (VII.18) has to be taken into account.
{\it Necessary and sufficient conditions of stability are obtained
 from (VII.20) when $\Delta^2 G$ couples pairs only}.
\bigskip

\noindent
{\it VII.5. Symmetric and Antisymmetric Single-Mode Perturbations }
\bigskip

In a single-mode analysis condition (VII.18) reduces to:
$$  m[  i \zeta_{lm}
\Sigma*_{lm} + m(|\psi_{lm}|^2 -|\eta_{lm}|^2)] +c.c. = 0 \eqno(VII.22)$$
For symmetrical and antisymmetrical modes, we take either real or imaginary
components
 of $\eta_{lm}$ and $ i \psi_{lm}$ in the spherical harmonic expansion, so the
Fourier
 expansion contains either $cos(m \varphi)$ or $sin(m \varphi)$. For such
perturbations,
 $\Delta^2 G$ becomes:
$$ \Delta^2 G = 4  \pi \Sigma_C \Omega a^4 \sum_{l,m
\ne 1,1} [\dot \Sigma_{lm} \zeta_{lm}^* ] + complex \  conjugate
\eqno(VII.23)$$
where we have taken into account the constraint (VII.17).
Notice that we have used without loss of generality we assume that
$\Sigma_{lm}$ is real (and so is
$\eta_{lm}$ and $i \psi_{lm}$), and equation (VII.23)
becomes:
$$ \Delta^2 G = 8  \pi \Sigma_C \Omega a^4  \sum_{l,m
\ne 1,1} [\dot \Sigma_{lm} \zeta_{lmR} ]. \eqno(VII.24)$$
We now have a pair coupling of $\Sigma_{lm}$ and  $\zeta_{lmR}$, hence we can
achive a
necessary and sufficient condition of stability. The imiginary part of
$\zeta_{lm}$
can now be obtained through equation (VII.22),
assuming non-radial modes (for radial modes equation (VII.22) is trivially
satisfied):
$$ \zeta_{lmI} = m {\psi_{lm}^2  - \eta_{lm}^2 \over \Sigma_{lm}}
\eqno(VII.25)$$
It is advantageous to introduce the following notations:
$$ z \equiv - k_{lm} { \eta_{lm} \over \Sigma_{lm}}  \qquad -\infty < z <
\infty
 \eqno(VII.26.a)$$
and
$$ x \equiv { m^2  \over k_{lm}^2} \qquad 0 \le x \le 1 . \eqno(VII.26.b)$$
For one single mode $(l,m)$, equation (VII.20) with equation (VII.25) reduces
to
in term of $z$:
$$Q < Q_{lm} = {(1-g_{lm}) \over P_4(z)} \eqno(VII.27.a)$$
we have assumed that $\zeta_{lmR}$ is equal to zero since any other value of
$\zeta_{lmR}$ will make $Q_{lm}$ bigger, and we are intrested only in its
lowest value.
$P_4(z)$ is a polynomial of order 4 in $z$,
$$P_4 (z) = 1 - {1 \over k_{lm}} \{({1 \over x} -1)[(1-x)z^4 +4(1-x)z^3 + 6z^2
+4z] +
 {1 \over x} \} \eqno(VII.27.b)$$
$P_4(z)$ has one and only one real maximum for any $x$.
We are only interested in values of $z$ for which $P_4(z)>0$. For $P_4(z)<0$,
$\Delta^2 V>0$ for any value of Q. The maximum of $P_4(z)$ is obtained for:
$$ z_{max}(x) = {x^{{1 \over 3}} \over \sqrt{1-x}} [(\sqrt{1-x}-1)^{1 \over 3}
+
 (\sqrt{1-x}+1)^{1 \over 3}] - 1 \eqno(VII.28.a)$$
for which
$$P_4 (z)_{max} = 1 - {1 \over k_{lm}} \{({1 \over x} -1)[(1-x)z_{max}^4 +
4(1-x)z_{max}^3 + 6z_{max}^2 +4z_{max}] + {1 \over x} \} \equiv 1 - {y(x) \over
k_{lm}}
 \eqno(VII.28.b)$$
 To any  pair of values $(l,m)$  corresponds a
 value $x$ defined by (VII.26.b) and a value $y(x)$. Following (VII.27.a) and
 (VII.28), we must thus  have
$$ Q<(Q_{lm})_{min} = {(1-g_{lm}) \over 1 - {y(x) \over k_{lm}}}
\eqno(VII.29)$$
The smallest minimum of $Q_{lm}$ is obtained for $(l,m)=(2,2)$ for which
 $ (Q_{22})_{min}={1 \over 2}$. Therefore the necessary and sufficient
condition of
stability with respect to symmetric or antisymmetric single mode perturbations
of
 Maclaurin disks is
$$ Q < {1 \over 2} \eqno(VII.30)$$
\bigskip

\noindent
{\it VII.6. Comments}
\bigskip

\noindent
a)  Binney and Tremaine (1987) have given the following dispersion relation for
the $\omega$ -
 modes of dynamical perturbations:
$$ \omega_r^3 - \omega_r \{ 4 \Omega^2 + k_{lm}[\Omega_0^2(1- g_{lm}) -
\Omega^2] \} +
2m\Omega[\Omega_0^2(1- g_{lm}) - \Omega^2]=0, \qquad \omega_r = \omega - m
\Omega
 \eqno(VII.31)$$
Stability holds if the non spurious $\omega_r$'s are real roots. Equation
(VII.31) is a
third order polynome. The condition for a polynome  of the form $x^3 + a_2 x
+a_3 = 0$
 to have only real roots is given in standard handbooks [for instance: Schaum's
 Mathematical Handbook (1968)]
$$ ({a_2 \over 3})^3+({a_3 \over 2})^2 < 0 \eqno(VII.32)$$
Inequality (VII.32) for equation  (VII.31) is exactly  our inequality (VII.29).
The results
of Binney \& Tremaine were obtained by solving three couple differential
equations,
our results were obtained without { \bf solving a single equation}.
\bigskip
\noindent
b) For radial modes, $m=0$, , $\Delta^2 J \equiv 0$ and
$$ \Delta^2 G = 4  \pi \Sigma_C \Omega a^4 \sum_{l=0} [\dot
\Sigma_{l0} \zeta_{l0} ]. \eqno(VII.33)$$
that is all radial modes are only pair coupled. Moreover, since [see equation
(VII.12.b)]
$$ \Sigma_{l0} = l(l+1) \eta_{l0} \eqno(VII.34)$$
 all $l$-modes are decoupled, $\Delta^2 V$ is a sum of squares of independent
modes.
 In this case one obtains stability limits for non single modes.
The most unfavorable limit, however, is again that given by $Q_{l0}$ for which
$y=4$
and $k_{ll} = l(l+1)$:
$$ Q <  Q_{l0} = {1 - g_{l0} \over 1 - {4 \over l(l+1)}} \qquad (l \ge 2),
\qquad g_{l0} = {l!^2 \over 2^{2l-1} [({l \over 2})!]^4} \eqno(VII.35)$$
\bigskip
\noindent
c) Notice the similarity of the mathematical analysis in the weak energy method
and the
strong method of paper II.
\bigskip

\noindent
{\bf VIII. Application to a Rayleigh's Shear Flow }
\bigskip

\noindent
{\it VIII.1. Stationary Configurations }

	The problem of the stability of a shear flow
 was considered by most developers of energy methods
 such as Arnold (1969), Holm at al. (1983) and
Grinfeld (1984). It interesting therefore to compare it to our own weak energy
method.
We consider a flow in a rectengular pipe, which occupies the domain:
$[0-L,0-1]$ that
is the rigid walls of the pipe are the lines $y=0, \ y=1$ while the lines $x=0,
\ x=L$
 are identified, in this way we construct a "torus". $x,y$ are the Caretesian
coordinates.
 We define the flow such that:
$$ \vec v_0 = W(y) \vec  1_x. \eqno(VIII.1)$$
The uniform density is set to unity
$$ \Sigma_0 = 1  \eqno(VIII.2)$$
Pressure and specific enthalpy are given by arbitary equations of state
$$ P =  P(\Sigma) \eqno(VIII.3.a)$$
$$ h = h(\Sigma) \eqno(VIII.3.b)$$
There is no potential of any kind in this problem.
Inserting equations (VIII.1-3) into the equations of motion (IV.7)
 we obtain the following:
$$ \Omega_c=0, \qquad b_{0y}=0 \eqno(VIII.4)$$
$b_{0x}$ is left undefined as was $\Omega_c$ in the models considered before.
 Notice in this problem we do not
have a "free boundary" but a fixed one, this modifies the possible
perturbations one
can construct such that:
$$ \delta \nu(0,y) = \delta \nu(L,y), \qquad \vec \xi(0,y) = \vec \xi(L,y)
\eqno(VIII.5.a)$$
$$ \xi_y(x,0) = \xi_y(x,1) = 0 \eqno(VIII.5.b)$$
since we consider here a torus with rigid boundaries.
\bigskip

\noindent
{\it VIII.2. Global Constraints }

 In this case $\Omega_c = 0 ,\  b_{0y}=0
$ are defined but not $b_{0x}$. This implies that not only do we have to fix:
$\Delta
\vec P = 0, \  \Delta J =0$ but $ \Delta^2 P_x = 0$ as well.
Inserting equations (VIII.1), (VIII.2) and (IV.5.b) in the first equation of
(IV.8.b) we obtain:
$$ \Delta \vec P = \int \Delta \vec v d^2 x = \int (- \vec \nabla \xi_x W +
\vec
\nabla \Delta \nu) d^2x = \vec 1_y \int - {\partial \xi_x \over \partial y} W
+
 {\partial \Delta \nu \over \partial y} d^2 x  = 0 \eqno(VIII.6.a)$$
Notice that the x-component which contained only boundary terms vanished do to
equations (VIII.5). The second order $ \Delta^2 P_x = 0$ becomes using equation
(IV.5.c)
$$\Delta^2 P_x = \int \Delta^2 v_x d^2 x = -2 \int {\partial \vec \xi \over
\partial
x} \cdot \Delta \vec v +  {\partial \Delta^2 \nu \over \partial x} d^2 x
 = -2 \int {\partial \vec \xi \over \partial
x} \cdot \Delta \vec v d^2 x \eqno(VIII.6.b)$$
$\Delta^2 \nu$ term vanishes due to integration by parts.
The angular momentum is fixed by:
$$ \Delta J = \vec 1_z \cdot \int \vec \xi \times \vec v_0 + \vec R \times
\Delta \vec
 v d^2 x = \int -\xi_y W + \vec 1_z \cdot( \vec R \times \Delta \vec v) d^2 x
\eqno(VIII.6.c)$$
\bigskip

\noindent
{\it VIII.3. $\Delta^2 V$ }

 We start from equation (IV.8.a). Using equations (A.6), (VIII.3) we have:
$$ \Delta (\vec \xi \cdot \vec \nabla  h) = \vec \xi \cdot  \vec \nabla \delta
h = \vec
\xi \cdot  \vec \nabla ({\partial h \over \partial \Sigma}|_0 \delta \Sigma) =
\vec
\xi \cdot  \vec \nabla (v_s^2 \delta \Sigma) \eqno(VIII.7)$$
we also used the definition of the velocity of sound $v_s$.  This means that
the
"potential" part of equation (IV.8.a) becomes using equations (VIII.5),
(IV.5.a) and
integrating by parts:
$$ \Delta^2 V_p = \int \Delta [\vec \xi \cdot \vec \nabla h ] d^2 x = \int
v_s^2
\delta \Sigma^2 d^2 x \eqno(VIII.7)$$
Using equation (IV.3) and (VIII.1) we have:
$$ \Delta \vec v = \delta \vec v + \xi_y W' \vec 1_x. \eqno(VIII.8)$$
Putting equations (VIII.8), (IV.5.c) into the "kinetic" part of equation
(IV.8.a) we have:
$$\Delta^2 V_k = \int [ \delta \vec v^2 + 2 \xi_y W' \delta v_x + \xi_y^2 W'^2
- 2 W
{\partial \vec \xi \over \partial x} \cdot (\delta \vec v + \xi_y W' \vec 1_x)]
d^2 x
\eqno(VIII.8)$$
$\Delta^2 \nu$ term vanishes due to integration by parts.
We use equation (IV.5.a) to subtitute $ {\partial \xi_x \over \partial x} =
-\delta
\Sigma - {\partial \xi_y \over \partial y}$ in equation (VIII.8) and obtain:
$$\Delta^2 V_k = \int [ \delta \vec v^2 + 2 \xi_y W' \delta v_x + \xi_y^2 W'^2
+ 2W (W'
\xi_y + \delta v_x) (\delta \Sigma + {\partial \xi_y \over \partial y}) - 2 W
{\partial
 \xi_y \over \partial x} \delta v_y] d^2 x  \eqno(VIII.9)$$
Integrating by parts the second and last terms and collecting some terms
together we
arrive at the following form:
$$  \eqalignno{ & \Delta^2 V_k = \int [\delta v_y^2 + (\delta v_x + W \delta
\Sigma)^2
 - W^2 \delta \Sigma^2
\cr & + 2 W \xi_y ({\partial \delta v_y \over \partial x} - {\partial \delta
v_x
 \over \partial y}) + 2 W W' \xi_y (\delta \Sigma + {\partial \xi_y \over
\partial y})
+ \xi_y^2 W'^2] d^2 x  &(VIII.10) \cr}$$
The term $\delta \omega = ({\partial \delta v_y \over \partial x} - {\partial
\delta v_x
 \over \partial y})$ which is the variation of the vorticity
can be written using equations (IV.5) and (VIII.1) as:
$$ \delta \omega = \xi_y W'' - \delta \Sigma  W' \eqno(VIII.11)$$
Inserting equation (VIII.11) into (VIII.10) and canceling the apropriate terms
we arrive
at the simpler expression:
$$ \Delta^2 V_k = \int [\delta v_y^2 + (\delta v_x + W \delta \Sigma)^2
 - W^2 \delta \Sigma^2 + W W'' \xi_y^2] d^2 x. \eqno(VIII.12)$$
Thus:
$$ \Delta^2 V = \Delta^2 V_k +  \Delta^2 V_p = \int [\delta v_y^2 + (\delta v_x
+ W
 \delta \Sigma)^2 + (v_s^2 - W^2) \delta \Sigma^2 + W W'' \xi_y^2] d^2
x.\eqno(VIII.13)$$
{}From this we extract the following famous sufficient condition:
$$ v_s^2 > W^2 \qquad and \qquad W W'' > 0. \eqno(VIII.14)$$
This was obtained by Lord Rayleigh (1880), Arnold (1966), Holm et al. (1983)
and
Grinfeld (1984). This is obiously a weak condition since we did not take full
advantage
of the constraints (VIII.6).
\bigskip

The author would like to thank J. Katz for critical discussions and for his
suggestion to
add section (III.1).

\vfill\eject

\noindent
{\bf References}

\refs
Arfken G. 1985 {\it Mathematical Methods for Physicists} Chap. 12 Academic
Press.
 P. 637 - 714.

Arnold V.I. 1966, {\it Journal de Mecanique} {\bf 5} 29.

Binney J. \& Tremaine S. 1987, {\it Galactic Dynamics} Chap. 5 Princeton
University Press.

Grinfeld M. 1984, {\it Geophys. Astrophys. Fluid Dynamics} {\bf 28} 31.

Holm D., Marsden J., Ratiu T. \& Weinstien A. 1983, {\it Physics Letters} {\bf
98A} 15.

Hunter C. 1963 {\it Month. Not. Roy. Astro. Soc.} {\bf 126} 24.

Katz J., Inagaki S. \& Yahalom A. 1993, {\it Publ. Astron. Soc. Japan} {\bf 45}
421.

Katz J. \& Lynden-Bell D. 1985, {\it Geophys. Astrophys. Fluid Dynamics}
{\bf 33} 1.

Lamb H. 1945 {\it Hydrodynamics} Dover Publications. P. 248

Lynden-Bell D. \& Katz J. 1981 {\it Proc. Roy. Soc. London} {\bf A 378}
179.

Rayleigh 1880 {\it Proc. London. Math. Soc.} {\bf 10} 4.

Spiegel M. 1968, {\it Mathematical Handbook} Macgraw-Hill, Schaum's outline
series
 p. 34.

Yahalom A., Katz J. \& Inagaki K. 1994, {\it Mon. Not. R. Astron. Soc.} {\bf
268} 506-516.

\endrefs

\vfill\eject
\noindent
{\bf  Appendix A: $ \Delta^2 V $ Variational Identity (V.8)}

We start from (IV.10). Inserting equation (V.1) into equation (IV.10) we
obtain:
$$\Delta^2 V = \int \{ (\Delta \vec v)^2
+(\vec \Omega \times \vec R) \cdot \Delta^2 \vec v
 + \Delta [\vec \xi \cdot \vec \nabla (h + \Phi)] \}|_0 \Sigma_0 d^2 x
\eqno(A.1)$$
Since we demand that $\Delta^2 J = 0$, we see from the second part of (IV.8.c)
that we can get read of the  $\Delta^2 \vec v$ term thus obtaining:
$$\Delta^2 V = \int \{ (\Delta \vec v)^2 -
 \vec \Omega \cdot (2 \vec \xi \times \Delta \vec v)
 + \Delta [\vec \xi \cdot \vec \nabla (h + \Phi)] \}|_0 \Sigma_0 d^2 x
\eqno(A.2)$$
The  terms containing $\Delta \vec v$ can be written as a difference of
squares:
$$(\Delta \vec v)^2  - \vec \Omega \cdot (2 \vec \xi \times \Delta \vec v)
 = (\Delta \vec v - \vec \Omega \times \vec \xi)^2 - (\vec \Omega \times \vec
\xi)^2
 = (\Delta \vec v - \vec \xi \cdot \vec \nabla \vec v_0)^2 - \Omega^2 \vec
\xi^2
 = (\delta \vec v )^2 - \Omega^2 \vec \xi^2 \eqno(A.3)$$
In the last equality we have used the second part of equation (IV.5.b).
 So equation  (A.2) with (A.3) can now be written as:
$$ \Delta^2 V = \int \{ (\delta \vec v)^2 - \Omega^2 \vec \xi^2
 + \Delta [\vec \xi \cdot \vec \nabla (h + \Phi)] \}|_0 \Sigma_0 d^2 x
\eqno(A.4)$$
Using the third operator identity of equation (IV.4), we may rewrite the
 $h + \Phi$ term in (A.1) as follows:
$$ \Delta [\vec \xi \cdot \vec \nabla (h + \Phi)] = \vec \xi \cdot \vec \nabla
\Delta
 (h + \Phi) - \vec \xi \cdot \vec \nabla \vec \xi \cdot \vec \nabla (h + \Phi)
 = \vec \xi \cdot \vec \nabla \delta (h + \Phi)
 + \vec \xi \cdot \vec \nabla  \vec \nabla (h + \Phi) \cdot \vec \xi
\eqno(A.6)$$
But from Euler's equations (IV.7.a) we see that
$$\vec \nabla (h + \Phi)|_0 = - \vec v_0 \cdot \vec \nabla \vec v_0
\eqno(A.7)$$
Subtituting (A.7) into (A.6) and $ \Delta [\vec \xi \cdot \vec \nabla (h +
\Phi)]$ back
 in (A.4), we find that
$\Omega^2 \vec \xi^2$ cancels out. Finally with $\delta \Sigma$ given in
(IV.5.a), and
 with some integration by parts, we obtain the following much simpler form for
 $\Delta^2 V$ which is that written in (V.8):
$$ \Delta^2 V = \int \{  \Sigma (\delta \vec v)^2
 + \delta \Sigma \delta (h + \Phi) \}|_0 d^2 x \eqno(A.8)$$
\bigskip

\noindent
{\bf  Appendix B: Spherical Harmonic Decomposition of $\Delta^2 J$}

Following the second part of (IV.8.c):
$$ \Delta^2 J = \vec 1_z \cdot \int (2 \vec \xi \times \Delta \vec v +\vec R
\times
 \Delta^2 \vec v)|_0  \Sigma_0 d^2 x. \eqno(B.1)$$
Inserting equation (IV.5.c) into (B.1) we obtain:
$$ \Delta^2 J = \vec 1_z \cdot \int (2 \vec \xi \times \Delta \vec v -2 \vec R
\times
  \vec \nabla \vec \xi \cdot \Delta\vec v )|_0  \Sigma_0 d^2 x. \eqno(B.2)$$
 Inserting (V.12.b) and (V.1) into (B.2) gives
$$  \eqalignno{ & \Delta^2 J = 2 a^4 \Omega \vec 1_z \cdot \int \{  [\vec
\nabla \eta +
 rot \vec \psi] \times [\vec \nabla (\zeta + \psi) + rot ( \eta \vec 1_z)] \cr
& -
\vec R \times  \vec \nabla  [\vec \nabla \eta + rot \vec \psi] \cdot
 [\vec \nabla (\zeta + \psi) + rot ( \eta \vec 1_z)] \} \Sigma_0 d^2 x. &(B.3)
\cr}$$
The spherical harmonic decomposition of $\Delta^2 J$ is obtained by replacing
in
 equation (B.3) $\zeta$ by (VII.16), $\Sigma_0$ by $\Sigma_C \chi$ (equation
(VII.1))
 and $\eta$, $\psi$ by (VII.5). Using also (VII.12.b) for $\Sigma_{lm}$, the
result comes
 out as follows:
$$\eqalignno{& \Delta^2 J = 4 \pi \Sigma_C a^4 \Omega \sum m[  i \zeta_{lm}
\Sigma*_{lm} + m(|\psi_{lm}|^2 -|\eta_{lm}|^2)] +c.c. &(B.4) \cr}$$
and demanding $ \Delta^2 J =0$ we obtain equation (VII.17).

\noindent
{\bf  Appendix C: Spherical Harmonic Decomposition of $\Delta^2 E$}

Let us start from (A.8) and take $\Delta^2 J = 0$. We shall set
$$ \Delta^2 V = \Delta^2 V_k + \Delta^2 V_p \eqno(C.1.a)$$
in which
$$ \Delta^2 V_k = \int  [\Sigma (\delta \vec v)^2]|_0 d^2 x \eqno(C.1.b)$$
and
$$ \Delta^2 V_p = \int [\delta \Sigma ( \delta h + \delta \Phi)]|_0 d^2 x .
 \eqno(C.1.c)$$
Inserting (V.12.a) in (C.1.b) gives:
$$ \Delta^2 V_k =  a^4 \Omega^2 \int \{ (\vec \nabla \zeta)^2 + 2 \vec \nabla
\cdot
 [\zeta rot (2 \eta \vec 1_z)] + [rot (2 \eta \vec 1_z)]^2 \} \Sigma_0 d^2 x
 \eqno(C.2)$$
Using Gauss's theorem, we can integrate (C.2) by part, and with
$\Sigma_0|_B=0$, obtain:
$$ \Delta^2 V_k = a^4 \Omega^2 \int [ - \zeta \vec \nabla \cdot ( \Sigma_0 \vec
\nabla
 \zeta) - 2 \zeta \vec \nabla \Sigma_0 \cdot rot (2 \eta \vec 1_z) - 4 \eta
\vec \nabla
 \cdot ( \Sigma_0 \vec \nabla \eta)] d^2 x.   \eqno(C.3)$$
The spherical harmonic decomposition of $\Delta^2 V_k$ is obtained by replacing
in
equation (C.3), $\zeta$ by (VII.16), $\Sigma_0$ by $\Sigma_C \chi$ (equation
(VII.1)) and
 $\eta$ by (VII.5.a). The result comes out as follows:
$$ \Delta^2 V_k = 4 \pi  \Sigma_C \Omega^2 a^4 \sum_{l=m, m=0}^{\infty}
[4 (k_{lm} - {m^2 \over k_{lm}}) |\eta_{lm}|^2 +  k_{lm} |\zeta_{lm} +{2 im
\over
k_{lm}} \eta_{lm}|^2 ] \eqno(C.4)$$
Notice that these are already some of the terms of $\Delta^2 V$ given in
(VII.18)

The spherical harmonic decomposition of $\Delta^2 V_p$ follows by inserting in
(C.1.c)
the respective expansions of $\delta \Sigma$ in (VII.12), $\delta h$ in
(VII.13) and
$\delta \Phi$ in (VII.15). For $ \Delta^2 V_p$,  we obtain :
$$ \Delta^2 V_p = 2 \pi a^4 \Omega_0^2 \sum_{l=m}^{\infty} \sum_{m=0}^{\infty}
         \Sigma_{lm}[\Phi^*_{lm}
	+ 3 \kappa \Sigma_C \Sigma^*_{lm}] + c.c. \eqno(C.5)$$
With (VII.15.b) and (VII.4), $\Delta^2 V_p$ can also be written:
$$ \Delta^2 V_p =  4 \pi \Sigma_C a^4 \Omega^2 \sum_{l=m, m=0}^{\infty}
 [-1 +  (1 - g_{lm}) {\Omega_0^2 \over \Omega^2}] |\Sigma_{lm}|^2  \eqno(C.6)$$
The sum of $\Delta^2 V_k$ given by (C.4) and $\Delta^2 E_p$ of (C.6) is the
expression
 of $\Delta^2 v$ written in (VII.18).

\bye